\documentstyle[epsf,12pt]{article}

\catcode`\@=11
\newbox\tempboxa
\newdimen\captionboxsubcount
\def\capsize#1{\captionboxsubcount=#1pt}
\newdimen\captionboxsub
\captionboxsub=\hsize \advance\captionboxsub by -\captionboxsubcount
\advance\captionboxsub by -\captionboxsubcount
\long\def\@makecaption#1#2{
 \setbox\@tempboxa\hbox{#1: #2}
 \ifdim \wd\@tempboxa >\captionboxsub
\rightskip=\captionboxsubcount \leftskip=\captionboxsubcount #1: #2
\else \hbox to\hsize{\hfil\box\@tempboxa\hfil}
 \fi}
\catcode`\@=12
\capsize{30}

\begin{document}

\begin{titlepage}
\begin{flushright}
KUNS-1354 \\
HE(TH) 95/13 \\
hep-ph/9509262 \\
\end{flushright}

\begin{center} \LARGE
     Dynamical Symmetry Breaking with\\
     Large Anomalous Dimension in Gauge Theories
\end{center}
\bigskip

\begin{center} \Large
        Yuhsuke Yoshida
\footnote{e-mail address :
{\tt yoshida@gauge.scphys.kyoto-u.ac.jp}} \\
\end{center}
\bigskip

\begin{center} \large \it
         Department of Physics, Kyoto University \\
         Kyoto 606-01, Japan
\end{center}

\begin{center} \Large \bf
Abstract
\end{center}

\begin{quote}
An analysis is given of the dynamical symmetry breaking of semi-simple
gauge groups.
We construct a class of renormalizable gauge theories for the
dynamically broken topcolor and technicolor interactions.
It is shown that a four-Fermi interaction in the strong coupling phase
emerges by the tumbling of semi-simple gauge groups in the low
energy region.
In our models the topcolor interaction provides the top quark with a
large anomalous dimension.
\end{quote}

\vspace{20pt}
\noindent
PACS numbers 11.15.Ex, 11.30.Qc, 11.30.Rd, 12.60.Fr, 12.60.Nz\,.

\end{titlepage}

\section{Introduction}
\label{sect:intro}

The dynamical symmetry breaking scenario of the Standard Model is a
fascinating issue.
Accordingly, the technicolor models\cite{simplestTC} and the top quark
condensation models\cite{TopCond} are considered.
However, there are theoretical and experimental difficulties in
many models.

The simplest technicolor models are excluded by the challenges of the
oblique corrections in the $W, Z$ gauge boson
self-energies,\cite{Oblique} and so are even for the walking
technicolor models.\cite{HarYos}
Then, the candidates for an acceptable technicolor model will have
spontaneously broken dynamics or have the techni-fermions with the
standard gauge symmetry invariant mass\cite{BFGH,Mae}.
However, in turn, we must trade naturalness for the vanishing
oblique corrections.
The flavor changing neutral current processes are also a
problem\cite{FCNC} to be overcome when we explain the masses of the
ordinary fermions by sideways mechanism\cite{ETC}.
We encounter the light pseudo Nambu-Goldstone (NG) bosons when we
use more than one doublet of techni-fermions.

The top condensation model\cite{TopCond} has severe problems of
naturalness and renormalizability, although the model can satisfy
all phenomenological constraints so far.
The phenomenological success is due to the dynamics providing a
large anomalous dimension $\gamma_m=2$ to the top quark bilinear
operator $\overline tt$.\cite{MTY}
When we formulate the model as a renormalizable gauge theory without
scalars, we are forced to introduce a strong coupling interaction,
such as technicolor, which dynamically breaks the topcolor gauge
symmetry.

Recently, a technicolor model assisted by the topcolor model was
proposed\cite{TCATC} in order to explain the large top quark mass and
the naturalness of the broken topcolor interactions.
In such a model the technicolor interactions are responsible for the
masses of the $W$ and $Z$ gauge bosons as well as the top-gluon.
The top quark mass is dynamically generated by the top quark
condensation and the masses of the other fermions are provided by
extended technicolor sideways.
However, many problems still remain unsolved.\cite{TCATCdiff}

In this paper, we show how to construct a class of topcolor assisted
technicolor models in the framework of the Schwinger-Dyson equation in
the improved ladder approximation.
We can also construct a renormalizable top quark condensation model.
Our theoretical models have the following properties; the
renormalizability, large top quark mass, the large anomalous dimension
$\gamma_m\simeq2$.

The top quark condensation model is based on the works in
Ref.~\cite{AF4F}.
It is shown that asymptotically free gauge theories with an additional
four-Fermi interaction has a non-trivial ultraviolet fixed point and
the large anomalous dimension within the (improved) ladder
approximation.
The present work is an extension of that work in part.
Our work is essentially based on that in Ref.~\cite{Luty}.
Semi-simple gauge groups are used for the tumbling gauge theory.
One gauge symmetry, which is a simple subgroup of the gauge group, is
broken by the gauge interaction of the other gauge symmetry.
We find the complete phase structure of the tumbling gauge theories with
semi-simple unitary gauge group.

This paper is organized as follows.
In section~\ref{sect:dsb} we study the dynamical symmetry breaking of
the semi-simple unitary group $SU(N_A)\times SU(N_B)$ in the framework of
the Schwinger-Dyson equation, and find the phase structure.
A system appears with an asymptotically free gauge interaction and a
four-Fermi interaction.
The detailed form of the coupled Schwinger-Dyson equation is given in
section~\ref{sect:CSDE}.
In section~\ref{sect:NG} we briefly show how the Nambu-Goldstone bosons
couple to the gauge currents.
The decay constants are given in terms of the fermion mass functions.
In section~\ref{sect:BDSCP} we solve a Schwinger-Dyson equation for
the top quark in the improved ladder approximation and show that the
top quark four-Fermi interaction appears in the strong coupling phase.

\section{Dynamical Symmetry Breaking in Gauge Theories}
\label{sect:dsb}

Although intuitive pictures \cite{Geo,RDS} of dynamical gauge symmetry
breaking are already given, there is an unsolved problem
especially in semi-simple gauge group\cite{Luty}.
What is the phase structure of such a system?
In this section we study the dynamical breaking of a semi-simple gauge
symmetry and solve the problem.

To begin with, we consider the semi-simple unitary gauge group
$G = SU(N_A)_A\times SU(N_B)_B$ for simplicity.
The gauge bosons of $SU(N_A)_A$ and $SU(N_B)_B$ are denoted by
$A^a_\mu$ and $B^a_\mu$, respectively.
It will be also interesting in general to consider anomaly safe groups
having complex representations such as $SO(10)$, $SO(14)$, $\cdots$
and $E_6$.
We introduce three kinds of fermions $\psi_R$, $\xi_L$ and $\eta_L$
transforming as $\psi_R\sim(\underline N_A,\underline N_B)$,
$\xi_L^i\sim(\underline N_A,1)$ and $\eta_L^j\sim(1,\underline N_B)$
for each $i$ ($=1,\cdots, N_B$) and $j$ ($=1,\cdots, N_A$) where
$\underline N$ represents the fundamental representation of the
unitary group $SU(N)$.
(see also Table.~\ref{tab:mycharge}.)
The fermions $\xi_L^i$ are $SU(N_B)_B$ singlets and the fermions
$\eta_L^j$ are $SU(N_A)_A$ singlets.
The subscripts $R$ and $L$ denote the usual chiral projections.
The gauge symmetry $G$ has no anomaly with this choice of matter
fields.
Then, the system consists of two gauge bosons and the three
types of fermions which minimally couple to the gauge bosons
according to their representations.
There is a global symmetry $SU(N_B)_\xi\times SU(N_A)_\eta$ acting on
these $\xi_L^i$ and $\eta_L^j$,
since fermions $\xi_L^1$, $\cdots$, $\xi_L^{N_A}$ are massless $N_B$
$\underline N_A$-plets and $\eta_L^1$, $\cdots$, $\eta_L^{N_B}$ are
massless $N_A$ $\underline N_B$-plets under $SU(N_A)_A$ and
$SU(N_B)_B$, respectively.
We may regard this global symmetry as a weak gauge symmetry by adding
the corresponding gauge bosons, which is irrelevant in the present
consideration of dynamical symmetry breaking.
The charge assignments of the fermions are summarized in
Table~\ref{tab:mycharge}.
\begin{table}[thbp]
\begin{center}
\begin{tabular}{c|cccc}
& $SU(N_A)_A$ & $SU(N_B)_B$ & $SU(N_B)_\xi$ & $SU(N_A)_\eta$ \\
\hline
$\psi_R$ & $N_A$ & $N_B$ & $1$   & $1$ \\
$\xi_L$  & $N_A$ & $1$   & $N_B$ & $1$ \\
$\eta_L$ & $1$   & $N_B$ & $1$   & $N_A$
\end{tabular}
\vspace{0.3cm}
\caption[]{
The charge assignments of the fermions.
}\label{tab:mycharge}
\end{center}
\end{table}

We first consider the extreme case where the $SU(N_B)_B$ gauge
coupling is turned off and only the $SU(N_A)_A$ gauge symmetry is
relevant.
We have a condensate $\langle\overline\xi_L\psi_R\rangle$
driven by the $SU(N_A)_A$ gauge interaction.
The most attractive channel is obvious in analogy with QCD.
The condensate $\langle\overline\xi_L\psi_R\rangle$ implies
that the $N_B$ pairs of two Weyl fermions $\xi_L^i$ and $\psi_R^i$
combine to form the $N_B$ massive Dirac fermions as
\begin{equation}
\Psi_A^i \equiv
\left(\begin{array}{c} \psi_R^i \\ \xi_L^i \end{array}\right) ~,
\end{equation}
where the superscript of $\psi_R^i$ is the index of the gauge group
$SU(N_B)_B$.
Owing to the custodial symmetry $SU(N_B)_B\times SU(N_B)_\xi$, the
condensate takes the form
$\langle\overline\Psi_{Ai}\Psi_A^j\rangle \propto \delta_i^j$ without
loss of generality.
This condensate breaks the symmetry $SU(N_B)_B$ completely, or more
precisely, breaks $SU(N_B)_B\times SU(N_B)_\xi$ down to the diagonal
subgroup $SU(N_B)_{B+\xi}$.
Accordingly the NG boson $\pi_A^a$ of the $SU(N_B)_{B+\xi}$ adjoint
representation appears, and the $SU(N_B)_B$ gauge boson becomes a
massive vector field of $SU(N_B)_{B+\xi}$ adjoint representation.
The same arguments hold for the opposite case where only the
$SU(N_B)_B$ gauge coupling is switched on.
In turn, the condensate $\langle\overline\eta_L\psi_R\rangle$
leads to the Dirac fermions
\begin{equation}
\Psi_B^j \equiv
\left(\begin{array}{c} \psi_R^j \\ \eta_L^j \end{array}\right) ~,
\end{equation}
where the superscript of $\psi_R^j$ is the index of the gauge group
$SU(N_A)_A$.
The condensate breaks the symmetry $SU(N_A)_A\times SU(N_A)_\eta$ down
to the diagonal subgroup $SU(N_A)_{A+\eta}$, and the NG boson
$\pi_B^a$ of the $SU(N_A)_{A+\eta}$ adjoint representation appears.

Now, let us consider the generic case in which both gauge couplings of
$G$ are turned on.
Although the physical picture is rather transparent\cite{Geo,RDS,Luty}
in analogy with the chiral symmetry breaking of QCD, the detailed
feature of dynamically breaking the gauge symmetry is complicated.
We have a possibility that both the gauge symmetries of $SU(N_A)_A$
and $SU(N_B)_B$ are dynamically broken by the condensates
$\langle\overline\eta_L\psi_R\rangle$ and
$\langle\overline\xi_L\psi_R\rangle$.
The resultant manifest symmetry is the global symmetry
$SU(N_A)_{A+\eta}\times SU(N_B)_{B+\xi}$.
This global symmetry is vector-like and cannot be broken because of
the Vafa-Witten theorem\cite{VafWit}.
As will be shown, the dynamical $SU(N_A)_A$ symmetry breaking is
solely caused by the (broken) $SU(N_B)_B$ gauge interaction, and
simultaneously the dynamical $SU(N_B)_B$ symmetry breaking is solely
caused by the (broken) $SU(N_A)_A$ gauge interaction.
Accompanied by the dynamical $SU(N_B)_B$ symmetry breaking, the NG
boson $\pi_A^a$ as well as the Dirac fermions $\Psi_A$ and
$\overline\Psi_A$ are formed by the $SU(N_A)_A$ gauge interaction.
This NG boson $\pi_A^a$ has a derivative coupling to the broken
$SU(N_B)_B$ current $J_B^{a\mu}$ with dimensionful coupling strength
$f_A$.
This quantity $f_A$ is the decay constant of $\pi_A^a$.
Owing to the minimal coupling, $g_B J_B^{a\mu} B_\mu^a$,
the NG boson $\pi_A^a$ is absorbed into the $SU(N_B)_B$ gauge boson
$B_\mu^a$ and this $B_\mu^a$ becomes massive vector field of
$SU(N_B)_{B+\eta}$ adjoint representation.
Here, we write the coupling constants of the $SU(N_A)_A$ and
$SU(N_B)_B$ gauge interactions as $g_A$ and $g_B$, respectively.
Simultaneously, the same argument holds for $\pi_B^a$, and the
$SU(N_A)_A$ gauge boson $A_\mu^a$ becomes massive.
We write the masses of $A_\mu^a$ and $B_\mu^a$ as $M_A$ and $M_B$,
respectively.
The masses, $M_A$ and $M_B$, are proportional to the decay constants,
$f_B$ and $f_A$, of the NG bosons, $\pi_B$ and $\pi_A$,\cite{JacJoh}:
\begin{eqnarray}
M_A^2 &=& g_A^2 f_B^2 ~,\nonumber \\
M_B^2 &=& g_B^2 f_A^2 ~,\label{eq:f-m relation}
\end{eqnarray}
respectively in the leading order of couplings.

On the other hand, the decay constants $f_A$ and $f_B$ depends on the
gauge boson masses $M_A$ and $M_B$, since the massive gauge bosons
$A_\mu^a$ and $B_\mu^a$ are responsible for forming the bound states
$\pi_A^a$ and $\pi_B^a$.
Namely, the gauge boson masses and the decay constants of the NG
bosons are consistently determined with each other.
It seems very complicated to study the dynamical symmetry breaking
systematically and quantitatively with the help of Schwinger-Dyson and
Bethe-Salpeter equations.
How do we disentangle the relation that output quantities $f_A$ and
$f_B$ are also input quantities $M_A$ and $M_B$?

Our key prescription for this problem is very simple.
We tentatively regard the gauge boson masses and the NG boson decay
constants as independent.
For given masses $M_A$ and $M_B$ we calculate the decay constant $f_A$
and $f_B$ using the SD and BS equations.
We vary the values of $M_A$ and $M_B$ as inputs.
Among the resulting sets $\{(M_A,M_B;f_A,f_B)\}$, we search for
the desired solution satisfying the relation (\ref{eq:f-m relation}).
We will explain a more systematic method later.

\begin{figure}[bhtp]
\begin{center}
\ \epsfbox{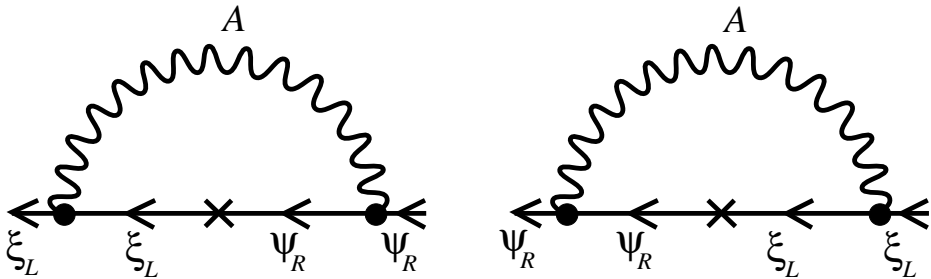} 
\caption[]{
The leading Feynman diagrams for the SD equation for $\Psi_A$
propagator.
Only the $SU(N_A)_A$ gauge boson $A^a_\mu$ contributes.
}
\label{fig:sdA}
\end{center}
\end{figure}

\begin{figure}[bhtp]
\begin{center}
\ \epsfbox{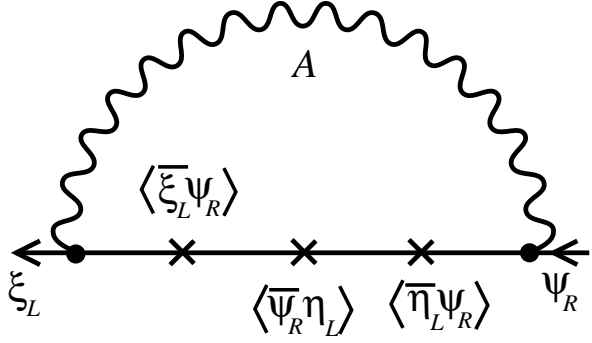} 
\caption[]{
The chiral transition $\langle\xi_L\overline\psi_R\rangle$ receives
a mixing effect by the condensates
$\langle\overline\eta_L\psi_R\rangle$ and
$\langle\overline\psi_R\eta_L\rangle$.
}
\label{fig:sdAB}
\end{center}
\end{figure}

Moreover, there are one further observation which makes the analysis
simpler.
As mentioned before, the condensates
$\langle\overline\xi_L\psi_R\rangle$ and
$\langle\overline\eta_L\psi_R\rangle$ are {\em solely} driven by the
gauge interactions $SU(N_A)_A$ and $SU(N_B)_B$, respectively.
For example, let us consider the propagator
$\langle\Psi_A\overline\Psi_A\rangle$.
The Weyl fermion $\xi_L$ is $SU(N_B)_B$ singlet, and the $SU(N_B)_B$
gauge boson $B^a_\mu$ does not interact with the $\xi_L$ component
of the Dirac fermion $\Psi_A$.
Then, the gauge boson $B_\mu^a$ cannot drive the fermions $\xi_L$ and
$\psi_R$ to make the chiral transitions $\xi_L \rightarrow \psi_R$ or
$\psi_R \rightarrow \xi_L$.
The chiral transitions $\xi_L \leftrightarrow \psi_R$ are
properly driven by the $A_\mu^a$ massive gauge boson.
The leading order terms of the Schwinger-Dyson equation for
$\langle\Psi_A\overline\Psi_A\rangle$ consist of the two diagrams
with only the $A_\mu^a$ massive gauge boson as in Fig.~\ref{fig:sdA}.
The similar argument holds for the propagator
$\langle\Psi_B\overline\Psi_B\rangle$ where the main contributions for
the chiral transitions $\eta_L \leftrightarrow \psi_R$ are given by
$B_\mu^a$.
More importantly the $\langle\xi_L\overline\psi_R\rangle$ propagator
receives mixing effects by the condensates
$\langle\overline\eta_L\psi_R\rangle$ and
$\langle\overline\psi_R\eta_L\rangle$.
The leading effect is depicted in Fig.~\ref{fig:sdAB}.
We take account of all such effects in the coupled Schwinger-Dyson
equations.

\begin{figure}[bhtp]
\begin{center}
\ \epsfbox{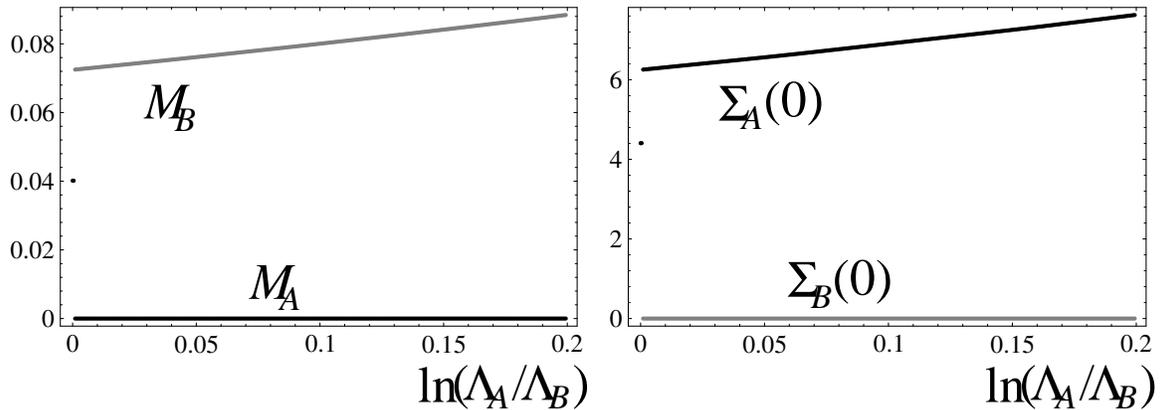} 
\caption[]{
The left hand side is the estimated gauge boson masses and the right
hand side is the mass functions $\Sigma_A(0)$ and $\Sigma_B(0)$ in the
case $N_A=N_B=3$.
The horizontal axes are the relative strength of the couplings
$\ln(\Lambda_A/\Lambda_B)$.
The black points indicate $M_A$, $\Sigma_A(0)$
and the gray points indicate $M_B$, $\Sigma_B(0)$.
A simple first order phase transition occurs at
$\ln(\Lambda_A/\Lambda_B) = 0$.
}
\label{fig:phase33}
\end{center}
\end{figure}

Let us study the phase diagram of the present system.
The coupled Schwinger-Dyson equations are easily solved by using a
numerical iteration (relaxation) method.
The detailed form will be given in section~\ref{sect:CSDE}.
The initial functional forms for the mass functions are taken as
symmetric; i.e., $\Sigma_A(x) = \Sigma_B(x)$.
When we evaluate the decay constants, we use a generalized
Pagels-Stokar formula which will be derived in section~\ref{sect:NG}.
The decay constants $f_A$ and $f_B$ are functions of the gauge boson
masses (and the interaction scales); $f_A=f_A(M_A,M_B)$,
$f_B=f_B(M_B,M_A)$.
Substituting this equations into Eqs.~(\ref{eq:f-m relation}), we find
that the gauge boson masses are determined by the intersection of the
following two equations
\begin{eqnarray}
M_A &=& g_Af_B(M_B,M_A) ~,\label{eq:MA}\\
M_B &=& g_Bf_A(M_A,M_B) ~.\label{eq:MB}
\end{eqnarray}
We can easily calculate the gauge boson masses numerically by applying
an iteration method to Eqs.~(\ref{eq:MA}) and (\ref{eq:MB}).
In order to make the analysis simple, we neglect the couplings
appearing in Eqs.~(\ref{eq:MA}) and (\ref{eq:MB}).
The values of $M_A$ and $M_B$ converge fast well.
The result is shown in Fig.~\ref{fig:phase33} with $N_A=N_B=3$.
We use a unit scale setting $\Lambda_B=1$ and fix the value of $N_B$
as $N_B=3$ below.
We observe three vacua at the point $\Lambda_A=\Lambda_B$.
It is seen, however, that the symmetric vacuum ($\Sigma_A=\Sigma_B$)
is unstable against the perturbation of the couplings.
If the symmetric vacuum was one of the stable points of the system,
we would have a plateau extending from the point $\Lambda_A=\Lambda_B$
to $\ln(\Lambda_A/\Lambda_B)>0$ in Fig.~\ref{fig:phase33}.
We conclude that the symmetric vacuum is an artifact generated by our
procedure and is not true vacuum.
Then, the correct solution shows a simple first order phase transition
at $\ln(\Lambda_A/\Lambda_B) = 0$.
Here, we notice that both the $SU(N_A)_A$ and $SU(N_B)_B$ broken
vacua are stable against any values of $\ln(\Lambda_A/\Lambda_B)$.
We recognize this fact by the explicit forms of the coupled
Schwinger-Dyson equations (in Eqs.~(\ref{eq:componentSD})).
For example, if we use asymmetric initial functions
($\Sigma_A(x) \ne 0$ and $\Sigma_B(x) \equiv 0$),
we will always find the $SU(N_B)_B$ broken vacuum having no dependence
of the values of the couplings.

As a result, we have the following phase.
In the range $-\infty<\ln(\Lambda_A/\Lambda_B)\le0$, the $SU(N_A)_A$
symmetry is broken with the gauge boson mass $M_A=f_B(0,M_A)$ and the
symmetry $SU(N_B)_B$ is completely manifest.
The values of the masses drastically change around the point
$\ln(\Lambda_A/\Lambda_B)=0$.
In the range $0\le\ln(\Lambda_A/\Lambda_B)<\infty$, $SU(N_A)_A$ is
completely manifest and $SU(N_B)_B$ is broken with $M_B=f_A(0,M_B)$.
This result shows that the most attractive channel (MAC) hypothesis
works completely.
The broken $SU(N_B)_B$ gauge interaction can form $SU(N_A)_A$ singlet
four-Fermi interaction by introducing $SU(N_A)_A$ singlet fermions.
(Notice that $\psi_R$ and $\xi_L$ cannot form such a four-Fermi
interaction.)

In conclusion, we obtain a theory in which {\it $SU(N_A)_A$ Yang-Mills
and $SU(N_B)_B$ four-Fermi interactions appear in the low energy
region} by tuning the couplings such that
$\ln(\Lambda_A/\Lambda_B)>0$.
There is no fine-tuning.
In certain appropriate regions the four-Fermi interaction is in the
strong coupling phase necessary for a dynamical symmetry breaking to
take place.
We will study this in section~\ref{sect:BDSCP}.

\begin{figure}[bhtp]
\begin{center}
\ \epsfbox{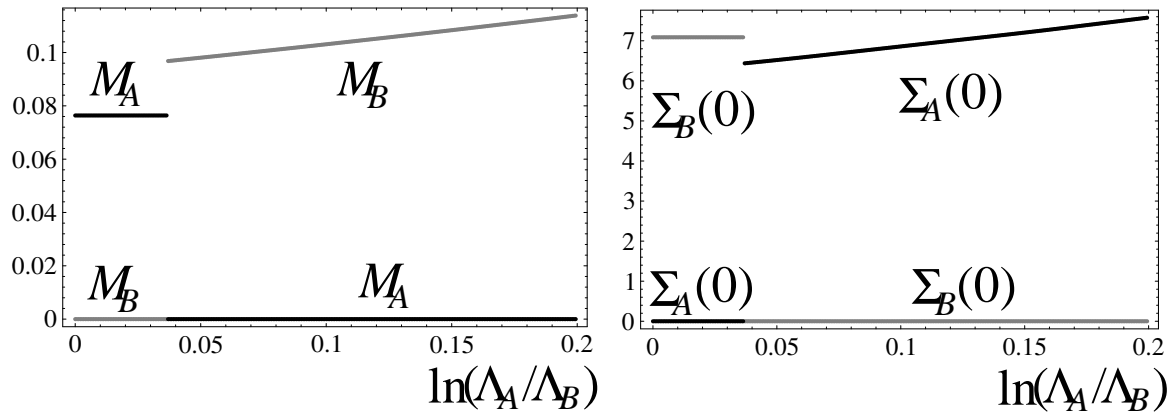}
\caption[]{
The phase diagram in the case $N_A=5$, $N_B=3$.
A first order phase transition occurs at $\ln(\Lambda_A/\Lambda_B) =
0.0355$.
}
\label{fig:phase53}
\end{center}
\end{figure}

\begin{figure}[bhtp]
\begin{center}
\ \epsfbox{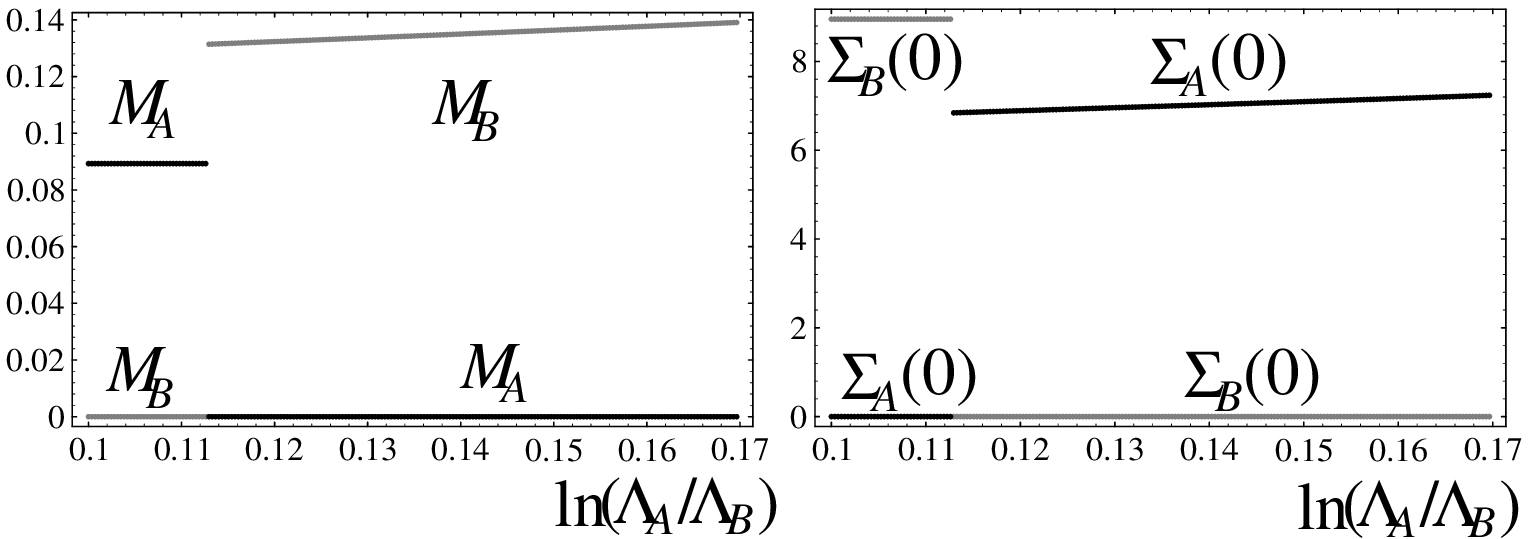}
\caption[]{
The phase diagram in the case $N_A=8$, $N_B=3$.
A first order phase transition occurs at $\ln(\Lambda_A/\Lambda_B) =
0.111$.
}
\label{fig:phase83}
\end{center}
\end{figure}

Next, let us take different gauge groups; i.e, $N_A \ne N_B$.
We fix the $SU(N_B)_B$ gauge group by setting $N_B=3$.
We change the value of $N_A$ as $N_A=5,8,9,10,12$.
As for the $N_A=5,8$ cases we have first order phase transitions.
The phase transition points move to the region
$\ln(\Lambda_A/\Lambda_B)>0$ as in Figs.~\ref{fig:phase53} and
\ref{fig:phase83}.

\begin{figure}[bhtp]
\begin{center}
\ \epsfbox{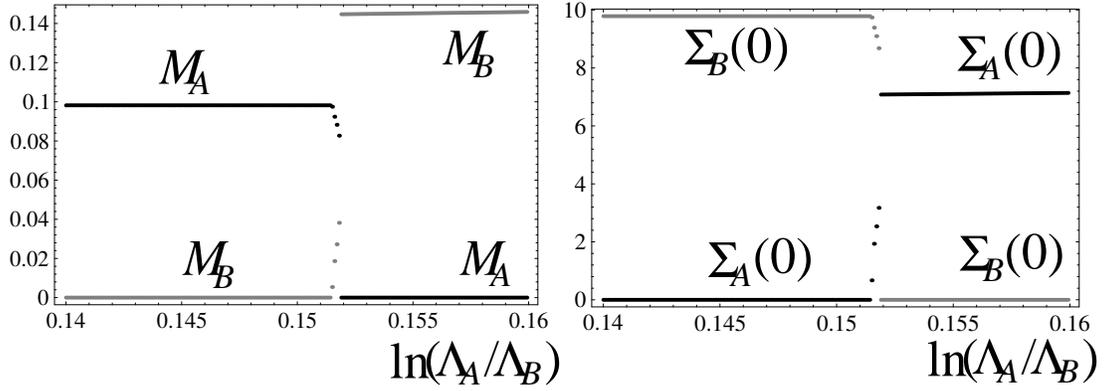}
\caption[]{
The phase diagram in the case $N_A=9$, $N_B=3$.
Phase transitions occur around $\ln(\Lambda_A/\Lambda_B) = 0.151
\sim 0.152$.
The former one is of second order and the latter one seems to be of
first order.
}
\label{fig:phase93}
\end{center}
\end{figure}

\begin{figure}[bhtp]
\begin{center}
\ \epsfbox{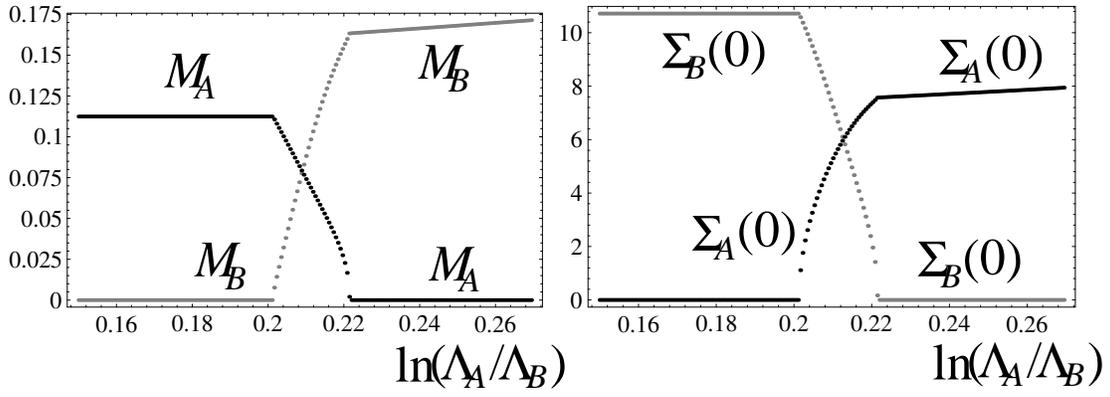}
\caption[]{
The phase diagram in the case $N_A=10$, $N_B=3$.
Second order phase transitions occur at
$\ln(\Lambda_A/\Lambda_B)=0.201, 0.222$.
}
\label{fig:phase103}
\end{center}
\end{figure}

\begin{figure}[bhtp]
\begin{center}
\ \epsfbox{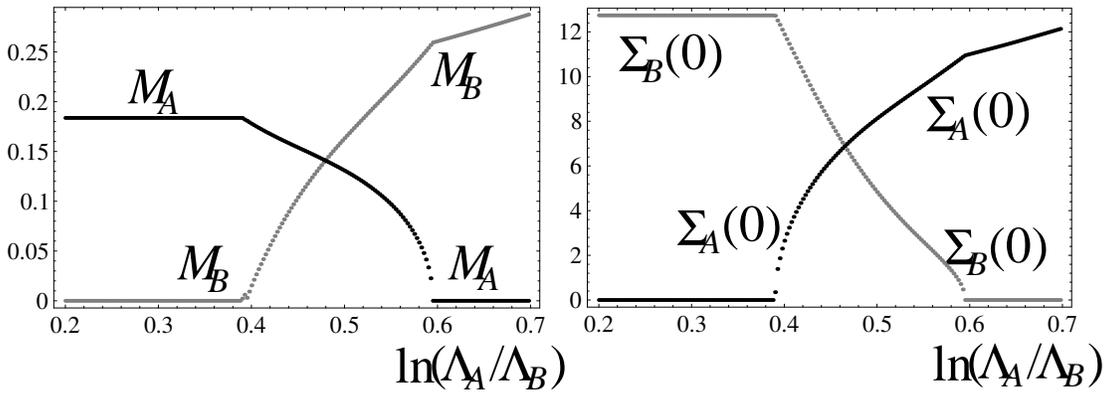}
\caption[]{
The phase diagram in the case $N_A=12$, $N_B=3$.
Second order phase transitions occur at
$\ln(\Lambda_A/\Lambda_B)=0.387, 0.597$.
}
\label{fig:phase123}
\end{center}
\end{figure}

However, for the $N_A=9$ case we have a second order phase transition
as in Fig.~\ref{fig:phase93} at $\ln(\Lambda_A/\Lambda_B)\cong0.1515$.
There are two phase transitions at $\ln(\Lambda_A/\Lambda_B) \cong
0.1515$ and $0.152$, and the latter one seems to be of first order.
Second order phase transitions occur clearly in the $N_A=10,12$
cases as in Figs.~\ref{fig:phase103} and \ref{fig:phase123}.
These models provide asymptotically free gauge theories with
additional strong coupling four-Fermi interaction around a second
order phase transition point.
This is studied in Ref.~\cite{AF4F}.

Here we note a fact.
If we use a initial condition such that $\Sigma_A(x) \equiv 0$ and
$\Sigma_B(x) \ne 0$, we always have a vacuum where the $SU(N_A)_A$
symmetry is broken and the $SU(N_B)_B$ symmetry is manifest.
Similar argument holds for the initial condition such that
$\Sigma_A(x) \equiv 0$ and $\Sigma_B(x) \ne 0$.
Namely, we can always have two different solutions specified by
$\Sigma_A(x)\equiv0$ and $\Sigma_B(x)\equiv0$, depending on the
initial functional forms of the mass function in solving the coupled
Schwinger-Dyson equations.
This property gives hysteresis curves in phase diagrams if we use
particular forms for the initial mass functions.

\section{The Coupled Schwinger-Dyson Equations}
\label{sect:CSDE}

In this section, we derive the coupled Schwinger-Dyson
equations used in the above analysis.
Before proceeding, we explain the basic ingredients in the
Schwinger-Dyson equations here.
The coupled Schwinger-Dyson equations with two massive gauge bosons
are specified by the eight parameters in the improved ladder
approximation: gauge boson masses $M_A$ and $M_B$, Yang-Mills
interaction scales $\Lambda=\Lambda_A,\Lambda_B$ of the running
couplings, one-loop $\beta$ functions
$\mu dg/d\mu = \beta(g) = - \beta_0 g^3$ for $g=g_A,g_B$, and the
second Casimir invariants $C_2(F)$ of the fermion fundamental
representations $F=\underline N_A,\underline N_B$.
Here, we are using the Yang-Mills interaction scale for specifying the
coupling strengths.
When we solve the coupled Schwinger-Dyson equations, one dimensionful
parameter out of $M_A$, $M_B$, $\Lambda_A$ and $\Lambda_B$ is
irrelevant.
We regard $\Lambda_B$ as a unit scale during the calculation by
rescaling all dimensionful parameters in terms of $\Lambda_B$.
In the present system the coefficient of the $\beta$ function and the
second Casimir invariant have different values in the two
Schwinger-Dyson equations.
Namely, $\beta_{A0} = (11N_A-2N_B)/(3\cdot16\pi^2)$,
$\beta_{B0} = (11N_B-2N_A)/(3\cdot16\pi^2)$ and
$T^a_XT^a_X = C_2(\underline N_X) = (N_X^2-1)/(2N_X)$ for $X=A,B$.

Now, let us write down the coupled Schwinger-Dyson equations.
The fermion propagator is defined by
\begin{equation}
S_F = \left(
\begin{array}{ccc}
\langle \psi_R\overline\psi_R \rangle &
\langle \psi_R\overline\xi_L  \rangle &
\langle \psi_R\overline\eta_L \rangle \\
\langle\xi_L \overline\psi_R \rangle &
\langle\xi_L \overline\xi_L \rangle  &
\langle\xi_L \overline\eta_L \rangle \\
\langle\eta_L \overline\psi_R \rangle &
\langle\eta_L \overline\xi_L \rangle  &
\langle\eta_L \overline\eta_L \rangle \end{array}
\right) ~.
\end{equation}
We note that the free fermion propagator is given by
\begin{equation}
S_{F0}(p) = \frac{i}{\alpha^\mu p_\mu} ~,
\end{equation}
where we define generalized $\sigma$ matrices as
\begin{equation}
\alpha^\mu = \left(\begin{array}{ccc}
\overline\sigma^\mu & 0 & 0 \\
0 & \sigma^\mu & 0 \\
0 & 0 & \sigma^\mu
\end{array}\right) ~,~~
\overline\alpha^\mu = \left(\begin{array}{ccc}
\sigma^\mu & 0 & 0 \\
0 & \overline\sigma^\mu & 0 \\
0 & 0 & \overline\sigma^\mu
\end{array}\right) ~,
\end{equation}
with
\begin{equation}
\overline\sigma^\mu = \sigma_\mu = \left(
\left(\begin{array}{cc} 1 & 0 \\ 0 & 1 \end{array}\right),
\left(\begin{array}{cc} 0 & 1 \\ 1 & 0 \end{array}\right),
\left(\begin{array}{cc} 0 & -i \\ i & 0 \end{array}\right),
\left(\begin{array}{cc} 1 & 0 \\ 0 & -1 \end{array}\right)
\right) ~.
\end{equation}
Then, the Schwinger-Dyson equation is given by
\begin{eqnarray}
iS_{F0}^{-1}(p)-iS_F(p)^{-1} &=& \int\frac{d^4k}{(2\pi)^4i}
\,g_A^2(-p^2\!-\!k^2)\,
\alpha^\mu T^a_A\, iS_F(k)\, \alpha^\nu T^a_A\, K_{\mu\nu}(p-k;M_A)
\nonumber\\
&+&
\int\frac{d^4k}{(2\pi)^4i} \,g_B^2(-p^2\!-\!k^2)\,
\alpha^\mu T^a_B \,iS_F(k)\, \alpha^\nu T^a_B\, K_{\mu\nu}(p-k;M_B) ~.
\nonumber\\ \label{eq:coupledSD}
\end{eqnarray}
The generators $T^a_A$ and $T^a_B$ eliminate $SU(N_A)_A$ and
$SU(N_B)_B$ singlet states, respectively.
The propagator $K_{\mu\nu}(l;M)$ of a massive gauge boson in a
Landau-like gauge\cite{MasNak} is given by
\begin{equation}
K_{\mu\nu}(l;M)
=
\frac{1}{M^2-l^2}
\left(g_{\mu\nu}+\frac{l_\mu l_\nu}{M^2-l^2}\right) ~.\label{eq:K}
\end{equation}
The quantity $g_X^2(\mu^2)$ ($X=A,B$) is the running coupling having a
threshold scale at the gauge boson mass $\mu=M_X$.
There is no need to regularize the running coupling $g_X^2$ below the
scale $\Lambda_X$ if $M_X>\Lambda_X$, since the running of $g_X^2$
stops below the scale $M_X$.
Then, $g_X^2$ takes the form
\begin{equation}
g_X^2(\mu^2)
= \frac{1}{\beta_{X0}\ln(\max(\mu^2,M_X^2)/\Lambda_X^2)}
 ~,\label{eq:g2 M>l}
\end{equation}
when we work in $M_X>\Lambda_X$.
In the cases with a massless gauge boson the running coupling should
be regularized, and various forms\cite{Higashi,ABKMN,KMY} may be
taken.
When we work in $M_X\le\Lambda_X$, we adopt the following
form\cite{ABKMN,HarYos}
\begin{equation}
g_X^2(\mu^2) =
   \frac{1}{\beta_{X0}} \times \left\{\begin{array}{ll}
\displaystyle \frac{1}{t} & \mbox{ if $t_F < t$ } \smallskip\\
\displaystyle \frac{1}{t_F} + \frac{(t_F - t_C)^2
   - (t - t_C)^2}{2t_F^2(t_F - t_C)} &\smallskip
   \mbox{ if $ t_C < t < t_F$ } \\
\displaystyle \frac{1}{t_F} + \frac{(t_F - t_C)}{2t_F^2} &
   \mbox{ if $ t < t_C$ } \smallskip
   \end{array}\right.~,
\label{eq:g2}
\end{equation}
where $X=A,B$, $t = \ln(\max(\mu^2,M_X^2)/\Lambda_X^2)$ and we fix
$t_F=0.15$ and $t_C=-2.0$.

Next, let us transform the SD equation in a component form.
We have vanishing condensates between the Weyl fermions with the same
chiralities:
\begin{equation}
\langle\overline\psi_R\psi_R \rangle =
\langle\overline\xi_L\xi_L \rangle =
\langle\overline\eta_L\eta_L \rangle =
\langle\overline\xi_L\eta_L \rangle =
\langle\overline\eta_L\xi_L \rangle \equiv 0 ~.
\end{equation}
Then, the non-trivial condensates are
$\langle\overline\psi_R\xi_L \rangle =
\langle\overline\xi_L\psi_R \rangle$ and
$\langle\overline\psi_R\eta_L \rangle =
\langle\overline\eta_L\psi_R \rangle$, where the condensates are taken
as real numbers by using phase transformations of the fermion fields.
Then, the mass function takes the form
\begin{equation}
\Sigma = \left(\begin{array}{ccc}
0 & 1_{B+\xi}\Sigma_A & 1_{A+\eta}\Sigma_B \\
1_{B+\xi}\Sigma_A & 0 & 0 \\
1_{A+\eta}\Sigma_B & 0 & 0\end{array} \right) ~,
\end{equation}
where $1_{B+\xi}$ and $1_{A+\eta}$ are $N_B\times N_B$ and $N_A\times
N_A$ unit matrices, respectively.
In the Landau-like gauge the wave function renromalizations are
expected to be small, then the fermion propagator is
given by
\begin{equation}
S_F(p) = \frac{i}{\alpha^\mu p_\mu - \Sigma(-p^2)} =
(\overline\alpha^\mu p_\mu + \Sigma(-p^2))\frac{-i}{\Sigma(-p^2)-p^2}
{}~.\label{eq:SF}
\end{equation}
Substituting Eq.~(\ref{eq:SF}) into Eq.~(\ref{eq:coupledSD}) and
carrying out the four-dimensional angle integrations, we find the
coupled Schwinger-Dyson equations in component form
\begin{eqnarray}
\Sigma_A(x) &=& \int_0^\infty\!ydy
K_A(x,y) \frac{\Sigma_A(y)}{y+\Sigma_A(y)^2+\Sigma_B(y)^2}
{}~,\nonumber \\
\Sigma_B(x) &=& \int_0^\infty\!ydy
K_B(x,y) \frac{\Sigma_B(y)}{y+\Sigma_A(y)^2+\Sigma_B(y)^2} ~,
\label{eq:componentSD}
\end{eqnarray}
where the kernel $K_X$ ($X=A,B$) is given by
\begin{equation}
K_X(x,y) =
\frac{ \lambda_X(x+y)}{x+y+M_X^2 +
\sqrt{(x+y+M_X^2)^2-4xy}}
\left(3+\frac{M_X^2}{\sqrt{(x+y+M_X^2)^2-4xy}}\right)
{}~,\label{eq:kernel}
\end{equation}
with
\begin{equation}
\lambda_X(x) \equiv \frac{C_2(\underline N_X)g_X^2(x)}{8\pi^2} ~.
\end{equation}

\section{Nambu-Goldstone Boson and its Decay Constant}
\label{sect:NG}

In this section we briefly show how the NG boson couples to the gauge
current with the decay constant and we derive a generalized
Pagels-Stokar formula in the present system.
In this paper instead of solving the BS equation for the NG boson
$\pi_A$ we use a convenient approximation by
Pagels and Stokar\cite{PagSto}, in which the BS amplitude is entirely
given by the mass function.
If we omit the interference effect of the other mass function, our
formula reduces to the usual Pagels-Stokar formula\cite{PagSto} up to
an overall factor.

In order for the argument to be transparent we concentrate on the
dynamical symmetry breaking of the gauge group $SU(N_B)_B$.
The Noether current of $SU(N_B)_B$ is given by
\begin{equation}
J_B^{b\mu} = \overline\psi_R\overline\sigma^\mu T_B^b \psi_R
+ \overline\eta_L\sigma^\mu T_B^b \eta_L ~,\label{eq:currentB}
\end{equation}
where $T_B^b$ is the generator of $SU(N_B)_B$.
The NG boson $\pi_A$ couples to the first part of this current
(\ref{eq:currentB}), and also couples to the left Weyl spinor current,
as
\begin{eqnarray}
\langle0\vert \overline\psi_R\overline\sigma^\mu T_B^b \psi_R
\vert \pi_A^a(q)\rangle &=&
i\delta^{ab}q^\mu~f_A \nonumber\\
\langle0\vert \overline\xi_L\sigma^\mu T_\xi^b \xi_L
\vert \pi_A^a(q)\rangle &=&
-i\delta^{ab}q^\mu~f_A ~.\label{eq:fA}
\end{eqnarray}
where $f_A$ is its decay constant.

The BS amplitude of the NG boson $\pi_A$ is defined by
\begin{equation}
\langle0\vert T \Psi_i(x/2)
\overline\Psi_j(-x/2) \vert\pi_A^a(q)\rangle
\equiv
1_A\; T_{B+\xi}^a\; \int\! dx e^{-ipx} \chi_{Aij}(p;q)
{}~,\label{eq:chiA}
\end{equation}
where $\Psi = (\psi_R, \xi_L, \eta_L)$ and $1_A$ is the
$N_A\times N_A$ unit matrix and denotes $SU(N_A)_A$ gauge singlet.
The truncated BS amplitude $\widehat\chi_A^a(p;q)$ is defined by
\begin{equation}
\widehat\chi_A^a(p;q) =
S_F^{-1}(p+q/2)\,\chi_A^a(p;q)\,S_F^{-1}(p-q/2) ~.
\end{equation}
Then, from Eqs.~(\ref{eq:fA}) and (\ref{eq:chiA}) the decay constant
$f_A$ is expressed in terms of the truncated BS amplitude
$\widehat\chi_A$ as
\begin{equation}
q^\mu~f_A = \frac{N}{2}
\int\! \frac{d^4p}{(2\pi)^4i}~{\rm tr}\;\overline\sigma^\mu\Big[
iS_F(p+q/2)\;\widehat\chi_A(p;q)\;iS_F(p-q/2)
\Big]_{11}~.\label{eq:fa=chi}
\end{equation}

The chiral Ward-Takahashi identity for the ``external'' symmetry
$SU(N_B)_B\times SU(N_B)_\xi$ is given by
\begin{equation}
-i q^\mu \Gamma_{A\mu}^a(p;q) = T_{B+\xi}^a\left(
S_F^{-1}(p+q/2)\gamma_5' - \gamma_5' S_F^{-1}(p-q/2) \right)
{}~,\label{eq:WTid}
\end{equation}
where $\gamma_5'= \mbox{diag.}(1,-1,0)$.
The NG boson $\pi_A^a$ couples to this vertex function as
\begin{equation}
\Gamma_{A\mu}^a(p;q) \sim -2f_A\,T_{B+\xi}^a\;
\widehat\chi(p;q)\;\frac{q_\mu}{q^2} + \cdots ~.\label{eq:NGpole}
\end{equation}
Using Eqs.~(\ref{eq:WTid}) and (\ref{eq:NGpole}), the truncated BS
amplitude is given in terms of the mass function in the soft momentum
limit $q_\mu\rightarrow 0$:
\begin{equation}
T_{B+\xi}^a\widehat\chi_A^a(p;0) =
\frac{T_{B+\xi}^a}{f_A}
\left(\begin{array}{ccc}
0 & -\Sigma_A(-p^2) & 0 \\
\Sigma_A(-p^2) & 0 & 0 \\
0 & 0 & 0
\end{array}\right) ~.\label{eq:chi=Sigma}
\end{equation}
In the Pagels-Stokar approximation we use the amputated BS amplitude
in the soft momentum limit instead of the full one.
Substituting Eq.~(\ref{eq:chi=Sigma}) into Eq.~(\ref{eq:fa=chi}) and
expanding the propagators in terms of $q_\mu$, we finally find
\begin{eqnarray}
\lefteqn{
f_A^2 = \frac{N_A}{16\pi^2}\int_0^\infty\!xdx\;
\Sigma_A(x)}\nonumber\\
&\times&\frac{\Sigma_A(x)\left(1+
\frac{\displaystyle 4x\Sigma_B'(x)\Sigma_B(x)-\Sigma_B(x)^2}
{\displaystyle 8x}\right)
- \frac{\displaystyle x}{\displaystyle2}\Sigma'_A(x)
\left(1+\frac{\displaystyle\Sigma_B(x)^2}{\displaystyle 2x}\right)}
{x+\Sigma_A(x)^2+\Sigma_B(x)^2} ~.
\end{eqnarray}

The above arguments similarly hold for the dynamical breaking of the
gauge symmetry $SU(N_A)_A$, and we obtain a similar formula for $f_B$.

\section{Broken Dynamics in the Strong Coupling Phase and the large
anomalous dimension $\gamma_m$}
\label{sect:BDSCP}

In section~\ref{sect:dsb} we find phase diagrams ($N_A\le9$) in which
a broken dynamics cannot break the other gauge symmetry.
In this section we study whether the broken dynamics is in a strong
coupling phase enough to break the $SU(2)_L$ symmetry.
For definiteness we consider the case where the symmetry $SU(N_A)_A$
is manifest and the symmetry $SU(N_B)_B$ is broken dynamically, and we
put $N_B=3$.
The topcolor gauge symmetry may be this $SU(3)_B$.
We consider the case when the $SU(2)_L$ gauge interaction is
relatively weak enough for us to regard it as a global symmetry.

Let us consider three Weyl fermions which are $SU(N_A)_A$ singlet.
The fermions have the following charge:
\begin{equation}
\begin{array}{c|cc}
    & SU(3)_B & SU(2)_L \\
\hline
q_L & 3 & 2 \\
t_R & 3 & 1 \\
b_R & 3 & 1
\end{array}
\end{equation}
The following analysis is devoted to make clear whether the condensates
$\langle\overline t_Rq_L\rangle=\langle\overline b_Rq_L\rangle$
form and break the $SU(2)_L$ symmetry.

The Schwinger-Dyson equation is simple and determines the propagator
of the fermions $q_L$ and $q_R$.
We use the notations $q_L^i\equiv(t_L,b_L)^T$ and
$q_R^i\equiv(t_R,b_R)^T$.
We use the improved ladder approximation and the Landau-like
gauge\cite{MasNak}.
The Dirac fermion $q$ respects the $SU(2)_{L+R}$ custodial symmetry,
and the propagator takes the form
\begin{equation}
\langle q^i(x) \overline q_j(0) \rangle
\equiv
\delta^i_j S_F(x) ~.
\end{equation}
The Dirac fermion propagator is expanded into two invariant amplitudes
as
\begin{equation}
iS_F^{-1}(p) = A(-p^2){p}\kern-6.0pt\mbox{\it/} - B(-p^2) ~.
\end{equation}
The Schwinger-Dyson equation determines the invariant amplitudes
$A(x)$ and $B(x)$, where $x\equiv -p^2$.

If we work with the massless gauge boson $M_B=0$, the amplitudes
$A(x)$ is identical to unity, which is shown after the four
dimensional angle integration.
Even when we work with $M_B\ne0$, it is verified in
Ref.~\cite{Kondo} that $A(x)\simeq1$ by an explicit numerical
calculation in the fixed coupling case.
In the high energy region $A(x)$ must converge to unity quickly
enough, otherwise the resultant Dirac fermion propagator will be
inconsistent with the result by the operator product expansion and the
renormalization group analysis.
It means an explicit breaking of the chiral gauge symmetries in the
present system.\cite{Higashi,CohGeo}
In this paper, we put $A(x)=1$ for simplicity although the coupling is
running.
It should not modify the physical consequences of this paper.

\begin{figure}[htbp]
\begin{center}
\ \epsfbox{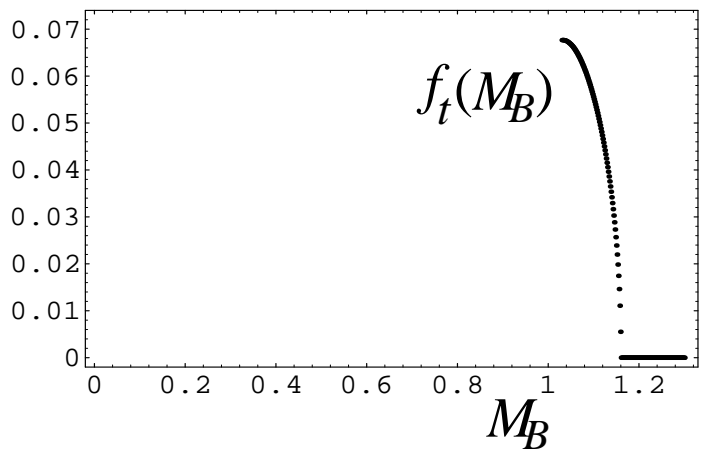} %
\caption[]{
The decay constant $f_t$ as a function of the gauge boson mass $M_B$.
Unit is $\Lambda_B$.
In the case $M_B>\Lambda_B$, the decay constant can be calculated
without regularizing the infrared form of the running coupling.
The phase transition occurs at $M_B=1.16\Lambda_B$ with second order.
}
\label{fig:notf}
\end{center}
\end{figure}

Then, the Schwinger-Dyson equation takes the form
\begin{equation}
\Sigma(x) = \int_0^\infty\!xdx\; K_B(x,y)
\frac{\Sigma(y)}{y+\Sigma(y)^2} ~.
\end{equation}
After obtaining the fermion propagator $S_F(p)$, we estimate the
decay constant $f_t$ of the NG boson $\pi_t^a$ by using the
Pagels-Stokar formula\cite{PagSto} which is given by
\begin{equation}
f_t^2 = \frac{3}{16\pi^2}\int_0^\infty\!xdx\;\Sigma(x)
\frac{\Sigma(x)-x\Sigma'(x)/2}{x+\Sigma(x)^2} ~.\label{eq:PS}
\end{equation}
We regard the decay constant as a function of $M_B$; i.e.,
$f_t=f_t(M_B)$.
We show the decay constant $f_t(M_B)$ in Fig.~\ref{fig:notf}.
The decay constant is evaluated with $M_B>\Lambda_B$, where we put
$\Lambda_B=1$.
This result has no ambiguity stemming from any regularized form of the
running coupling in the infrared region lower than the interaction
scale $\Lambda_B$.
This is in contrast with the result in Ref.~\cite{ABKMN}.
Above the critical mass $M_B>f_t^{-1}(0)=1.16\Lambda_B$ the chiral
symmetry is restored.
Near the phase transition point ($M<f_t^{-1}(0)$) with a small
dynamical mass of the Dirac fermion, the anomalous dimension
$\gamma_m$ approaches that of the Nambu-Jona-Lasinio model:
\begin{equation}
\gamma_m \equiv
- \frac{d}{d\ln \mu}\ln\vert\langle\overline q q\rangle\vert
\cong \frac{d}{d\ln M_B}\ln\vert\langle\overline q q\rangle\vert
= 2 ~,
\end{equation}
in the low energy region $\mu <\!< M_B$.
Here the gauge boson mass $M_B$ plays the role of the cutoff.
We conclude that {\em the renormalizable theory, considered here,
provides the large anomalous dimension $\gamma_m\cong 2$ just below
the critical point $M_B\cong f_t^{-1}(0)$ in the low energy region}.

Next, let us proceed to the lower region $M_B<\Lambda_B$.
We can guess that $f_t(M_B)$ will be a constant in $M_B<\Lambda_B$,
since such a mass $M_B$ smaller than $\Lambda_B$ should be negligible
in the gauge interaction dynamics.
This is also confirmed empirically.
If we convert the value $f_t(\Lambda_B)/\Lambda_B$ to the ordinary QCD
case by multiplying by a necessary factor, it already saturates the
experimental value $93~[\mbox{MeV}]/\Lambda_{QCD}$ of the actual pion.
Therefore, we should keep the property $f_t(M_B) \sim \mbox{const.}$
($M_B<\Lambda_B$) when we regularize the running coupling in the
infrared region.
\begin{figure}[htbp]
\begin{center}
\ \epsfbox{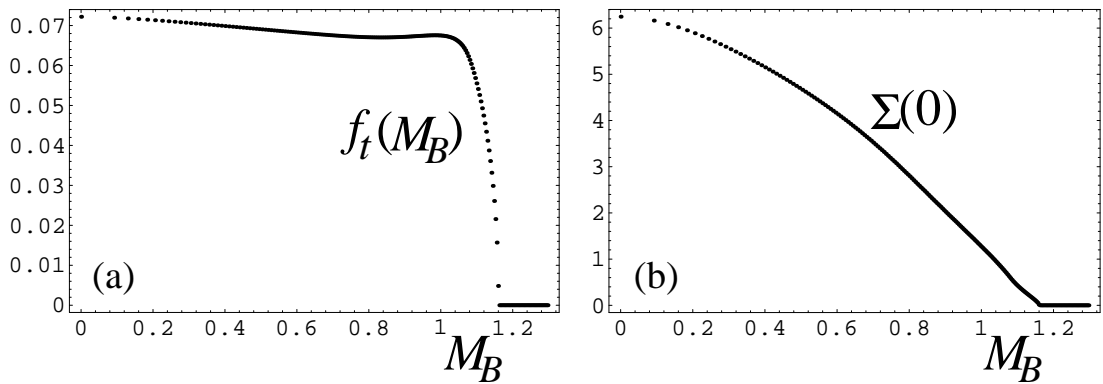}
\caption[]{
The decay constant $f_t$ is plotted as a function of the gauge boson
mass $M_B$ in Fig.~(a).
The mass function of the top quark is plotted in Fig.~(b).
The unit is in $\Lambda_B$.
As mentioned in Fig.~\ref{fig:notf} the phase transition occurs at
$M_B=1.16\Lambda_B$.
}
\label{fig:tf15}
\end{center}
\end{figure}
We adopt the form (\ref{eq:g2}).
Of course we should use the same coupling as that used in
Eq.~(\ref{eq:coupledSD}).
Above the scale $\mu^2=\exp t_F$ the functional form is exactly the
same as that of the one-loop running coupling, below the scale
$\mu^2=\exp t_F$ the running coupling is regularized by using the
second order polynomial in $t=\ln\mu^2$ and in the low energy region
$t\le t_C$ the coupling becomes a constant.
The running coupling with $t_F=0.15$ agrees with the one-loop running
coupling form over almost of the range of $t$ greater than the
interaction scale $\Lambda_B$.
The smaller value of $t_F$ would be good but increases the error in
numerical calculations.
The result of $f_t(M_B)$ is shown in Fig.~\ref{fig:tf15}.
We should notice that the functional form above
$M_B^2\ge\exp t_F \simeq 1.16$ in Fig.~\ref{fig:tf15} is exactly the
same as that in Fig.~\ref{fig:notf}, and is perfectly independent on
the infrared regularization of the running coupling, since the running
of the coupling stops below the threshold $M_B^2$ which is just above
the regularized scale $\exp t_F$.
We observe that $f_t(0) = 0.0722$ and $f_t(\Lambda_B) = 0.0675$.
So, the decay constant squared $f_t$ does not change by more than
7\% in the range $0<M_B<\Lambda_B$.

\clearpage
\begin{center}
\Large Acknowledgements
\end{center}
We would like to thank T.~Kugo for enlightening suggestion,
discussions and comments.
We are grateful to M.E.~Peskin for discussions and suggestions.
We are also grateful to K.~Higashijima and T.~Maskawa for comments.
We would like to thank A.~Bordner for correcting English.
We also thank Y.~Kikukawa, N.~Maekawa and K.~Nakanishi for
discussions.

\end{document}